\documentclass[preprint2]{emulateapj}

\newcommand{\fer}{{\it Fermi}}
\newcommand{\swf}{{\it SWIFT}}
\newcommand{\wse}{{\it WISE}}
\newcommand{\bzcat}{{\it Roma-BZCAT}}
\usepackage{graphicx}
\usepackage{longtable}

\slugcomment{pre-proof version}

\shorttitle{Radio Weak BL Lac objects in the \fer\ era}
\shortauthors{F. Massaro, E. Marchesini, R. D'Abrusco, N. Masetti, I. Andruchow \& Howard A. Smith 2016}

\begin{document}
\title{Radio Weak BL Lac objects in the \fer\ era}
\author{F. Massaro\altaffilmark{1,2,3}, E. J. Marchesini\altaffilmark{1,4,5}, R. D'Abrusco\altaffilmark{6}, N. Masetti\altaffilmark{7,8}, I. Andruchow\altaffilmark{4,5} \& Howard A. Smith\altaffilmark{6}}

\affil{Dipartimento di Fisica, Universit\`a degli Studi di Torino (UniTO), via Pietro Giuria 1, I-10125 Torino, Italy.}
\affil{Istituto Nazionale di Fisica Nucleare, Sezione di Torino, via Pietro Giuria 1, I-10125 Torino, Italy.}
\affil{INAF--Osservatorio Astrofisico di Torino, via Osservatorio 20, 10025, Pino Torinese, Italy.}
\affil{Facultad de Ciencias Astron\'omicas y Geof\'isicas, Universidad Nacional de La Plata, Paseo del Bosque, B1900FWA, La Plata, Argentina.}
\affil{Instituto de Astrof\'isica de La Plata, CONICET--UNLP, CCT La Plata, Paseo del Bosque, B1900FWA, La Plata, Argentina.}
\affil{Smithsonian Astrophysical Observatory, 60 Garden Street, 02138 Cambridge (MA), United States.}
\affil{INAF--Istituto di Astrofisica Spaziale e Fisica Cosmica di Bologna, via Gobetti 101, I-40129, Bologna, Italia.}
\affil{Departamento de Ciencias F\'isicas, Universidad Andr\`es Bello, Fern\`andez Concha 700, Las Condes, Santiago, Chile}

\begin{abstract}
The existence of ``radio weak BL Lac objects'' (RWBLs) has been an open question, still unsolved, since the discovery that quasars could be radio-quiet or radio-loud. Recently several groups identified RWBL candidates, mostly found while searching for low energy counterparts of the unidentified/unassociated gamma-ray sources listed in the \fer\ catalogs. Confirming RWBLs is a challenging task since they could be confused with white dwarfs or weak emission line quasars when there are not sufficient data to precisely draw their broad band spectral energy distribution and their classification is mainly based on a featureless optical spectra. Motivated by the recent discovery that \fer\ BL Lacs appear to have very peculiar mid-IR emission, we show that it is possible to distinguish between WDs, WELQs and BL Lacs using the [3.4]-[4.6]-[12]$\mu$m color-color plot built using the \wse\ magnitudes when the optical spectrum is available. On the basis of this analysis, we identify WISE J064459.38+603131 and WISE J141046.00+740511.2 as the first two genuine RWBLs, both potentially associated with \fer\ sources. Finally, to strengthen our identification of these objects as true RWBLs, we present multifrequency observations for these two candidates to show that their spectral behavior is indeed consistent with those of the BL Lac population.
\end{abstract}


\section{Introduction}
\label{sec:intro}
BL Lac objects constitute the most elusive class of radio-loud active galactic nuclei \citep[AGNs;][]{urry95}. Hosted in elliptical galaxies \citep[see e.g.,][]{falomo14}, their emission extends from radio frequencies up to the highest TeV energies and is characterized by rapid and large-amplitude flux variability, flat radio spectra, even below $\sim$1GHz \citep[see e.g.,][]{massaro13a,massaro13b}, peculiar infrared (IR) colors \citep{paper1,paper2} and significant and variable optical polarization (i.e., $P\sim10\%$) \citep[see e.g.,][and references therein]{andruchow05,smith07}. 

BL Lacs show featureless optical spectra, sometimes having weak emission lines of rest-frame equivalent width ($EW_r$) lower than 5\AA\ or absorption lines due to their host galaxies, both superimposed to a blue continuum \citep[see e.g.,][]{laurent99}. BL Lac objects are one of the two subclasses of the radio-loud AGNs known as blazars \citep[see e.g.,][for a recent review]{massaro09}. The other subclass is constituted by the flat spectrum radio quasars (FSRQs), characterized by properties similar to those of BL Lacs but having a quasar-like optical spectra \citep[i.e., with broad emission lines; see e.g.,][]{veron00} and radio spectral indices\footnote{Spectral indices, $\alpha$, are defined by flux density, S$_{\nu}\propto\nu^{-\alpha}$} lower than 0.5 between $\sim$300 MHz and a few GHz \citep[see][and references therein]{massaro14a}. 

Over the entire electromagnetic spectrum both BL Lacs and FSRQs feature a typical double-bumped spectral energy distribution (SEDs) with the first component peaking between the IR and the X-rays and the second one in the gamma-rays \citep[see e.g.,][]{padovani95,paggi09,abdo10a}. Mid-IR colors clearly helps to determine the energy peak location of the low energy component \citep{paper1}. BL Lac SEDs and their broad-band features were described in terms of emission arising from particles accelerated in relativistic jets directly pointing towards the observer \citep{blandford78}.

BL Lacs are traditionally discovered using radio, X-ray surveys and/or their combination \citep[see e.g.,][]{stickel91,sambruna96,giommi05}, since in the optical energy range the lack of spectral features combined with their flux variability makes a color-color selection extremely challenging \citep{jannuzi93,londish02,collinge05,plotkin08}.  However, recent observations to search for optical polarization in BL Lacs selected using deep optical survey as the Sloan Digital Sky Survey \citep[see e.g., SDSS DR12][and references therein]{alam15} and the Two-Degree Field Quasar Redshift Survey \citep[2QZ][and references therein]{smith05} validated this strategy of investigating proper motion and broad-band color properties to minimize contamination by featureless Galactic objects while searching for quasi-featureless spectra \citep{smith07}. 

At the current moment, the most efficient survey for discovering new BL Lacs is the \fer\ gamma-ray one \citep{abdo10b,ackermann11a,nolan12}, in particular when combining gamma-ray observations with mid-IR colors analyses \citep{massaro12a,massaro12b} and follow up optical spectroscopic observations \citep{paggi13,shaw13,massaro15,refined}. This occurs because BL Lacs are the largest known population emitting at MeV-GeV energies, constituting more than 30\% of the associated sources in the latest release of the \fer\ catalog \citep[The \fer\ - Large Area Telescope Third Source Catalog, 3FGL][]{acero15}. Moreover, recent optical spectroscopic campaigns aimed to confirm the nature of blazar candidates of uncertain type listed in the 3FGL showed that a significant fraction of those observed to date are also BL Lacs \citep[see e.g.][]{crespo16a,crespo16b,crespo16c}.

However, nowadays, all the known \fer\ BL Lacs have radio counterparts and radio surveys are therefore crucial to associate $\gamma$-ray objects with their counterparts. Thus, one of the major open issues for this rare AGN class is proving the existence of radio-weak BL Lac objects (RWBLs). We indicate them as radio-weak sources instead of using the term {\it ``radio-quiet''}, generally adopted for the AGN unification scenario. This designation is more useful because radio follow up observations could reveal some radio emission. However, the current lack of counterparts in the major radio surveys available to date place them in the tail of the distribution of the IR-to-radio flux ratios for the whole BL Lac population.

At the current moment only a handful of sources have been claimed to be RWBLs on the basis of their broad band properties as featureless optical spectra combined with X-ray emission and and their lying within the positional uncertainty regions of unidentified/unassociated gamma-ray sources \citep[UGSs;][]{paggi13,ricci15,landoni16}. 

Their existence will have a crucial impact (i) on the estimates of their source counts and luminosity functions \citep{ajello12,ajello14}, a task already complicated by the challenge of measuring their redshifts \citep{falomo14}, (ii) on the gamma-ray association methods all currently based on radio surveys \citep[see e.g,][and references therein]{ackermann15}, if proved that RWBLs could be associated to \fer\ sources, as well as (iii) on the unification scenario of AGNs \citep{urry95}, to name a few implications. An accurate accounting is crucial to compute correctly their contribution to the extragalactic gamma-ray background for which, together with FSRQs \citep{ajello12} and radio galaxies \citep{inoue11,dimauro14,lobes} they provide the most significant ones. Additional implications are related to jet physics. Radio emission in extragalactic radio-loud AGNs permits us to investigate the ``large-scale'' (i.e. from pc to Mpc size) evolution of relativistic jets and their interactions with the environments \citep[see e.g.,][and references therein]{liuzzo13,nulsen13,fabian12}. The existence of RWBLs could be crucial to address different phases of jet formation in cases when relativistic jets are not able to form extended radio structures \citep{blandford79,begelman84}, as it seems from low radio frequency (i.e., below $\sim$1 GHz) spectral studies \citep[see][and references therein]{massaro13b,ugs6lowfrq,mwabl}.

There are two major problems to solve before claiming the existence of RWBLs, even when featureless optical spectra are available. They can be confused with or misinterpreted as hot white dwarfs \citep[WDs; see e.g.][]{kepler15} and/or weak emission line quasars \citep[WELQs;][and references therein]{ds9,plotkin10} when optical spectroscopic data are available. { Optical polarimetric observations could shed light on their nature. However, unfortunately, optical polarization in BL Lacs can be variable and it is not always observed thus not representing the most definitive stamp to confirm their nature \citep[see e.g.,][]{londish04}.}

In the present paper we explore the use mid-IR colors as a diagnostic tool to disentangle between BL Lacs, WDs and WELQs. This new tool could potentially open a new window to search for RWBLs. This investigation is motivated by the recent discovery that \fer\ BL Lacs show peculiar mid-IR colors \citep[see also,][and references therein]{gir}, result made possible by observations performed with the \wse\ all-sky survey \citep{wright10}, covering a region well isolated from that of other Galactic and extragalactic objects. The main explanation underlying this distinction is the BL Lac non-thermal IR emission \citep[see e.g.,][]{paper2,ugs1}. 

This manuscript is organized as follows. In \S~\ref{sec:samples} we present all the samples used to perform our mid-IR color-color analysis. In \S~\ref{sec:analysis} we compare the mid-IR colors of BL Lacs with those of WELQs and WDs, while in \S~\ref{sec:rwbls} we search for RWBLs. \S~\ref{sec:variability} is dedicated to the analysis of the IR variability for the best candidates selected on the basis of the mid-IR colors ad the optical spectra. Then in \S~\ref{sec:swift} we also investigate the \swf\ observations available for these candidates while in \S~\ref{sec:seds} we discuss on the SEDs. Summary and conclusions are given in \S~\ref{sec:summary}

Unless stated otherwise, for our numerical results, we use cgs units. \wse\ magnitudes used here are in the Vega system and are not corrected for the Galactic extinction. As shown in our previous analyses, such corrections mostly affect the magnitude at 3.4$\mu$, for sources lying at low Galactic latitudes and it ranges only between 2\% and 5\% \citep[see e.g.][]{wibrals}, thus being negligible. In addition this allows us to compare plots in the present analysis with those of previous investigations \citep{ugs2}.

\section{Sample selection}
\label{sec:samples}
To perform our analysis we built three main samples. The first two composed of WDs and WELQs, respectively; both used in comparison with the third one of \fer\ blazars. We searched for mid-IR counterparts of all samples in the latest release of the \wse\ catalog\footnote{http://wise2.ipac.caltech.edu/docs/release/allwise/}. All our final samples of WDs, WELQs and gamma-ray blazars only include sources having a \wse\ counterpart within 3\arcsec.3 that is detected at 3.4$\mu$m, 4.6$\mu$m and 12$\mu$m so allowing us to build at least one of the mid-IR color-color plots used in our previous blazar analyses \citep{wibrals,gir}. The choice of 3\arcsec.3 association radius is the same adopted in our previous investigations \citep{ugs1} and it was computed on the basis of Montecarlo procedure corresponding to a probability of having spurious associations of the order of $\sim$1\% \citep[see also][for additional details]{wibrals}.

\subsection{BL Lac objects and Flat Spectrum Radio Quasars}
The first sample used in our analysis lists all the blazars, both BL Lacs and FSRQs, included in the \bzcat\footnote{http://www.asdc.asi.it/bzcat/} \citep{massaro15c} and having a counterpart in the latest release of the \fer\ catalog \citep{acero15}. This sample has been recently used to perform the analysis on the {\it IR--gamma-ray connection} \citep{gir}. According to the nomenclature proposed in the \bzcat\ hereinafter we will refer to BL Lac objects as BZBs and to the FSRQs as BZQs. We choose this sample because it has the largest fraction of BZBs with a mid-IR counterpart in \wse\ corresponding to 603 out of 610 (i.e., $\sim$99\%) within our association radius; this fraction is similar to that occurring for the BZQs with 419 out of 426 (i.e., $\sim$98\%).

\subsection{White Dwarfs}
Hot WDs could be misclassified as BZBs because they could show almost featureless optical spectra and X-ray radiation \citep[see e.g.,][]{heise85,bilikova10}. Thus in the present analysis we want to show that the mid-IR colors of the \wse\ counterparts of WDs is different from that of BZBs. We expect that for WDs mid-IR emission arises from the presence of a companion brown dwarf or the presence of a dusty disk\footnote{http://www.stsci.edu/$\sim$debes/wired.html}, both expected to have thermal IR colors \citep{debes11} while those of BZBs are dominated by the non-thermal synchrotron radiation of particles accelerated in their jets.

To achieve this goal we considered Sloan Digital Sky Survey DR7 White Dwarf Catalog (SDSSDR7WD) \citep{kleiman13}. This new catalog of spectroscopically confirmed WDs from the Sloan Digital Sky Survey (SDSS) Data Release 7 (DR7) listing more than 19000 stars. Crossmatching the SDSSDR7WD catalog with the latest release of the \wse\ catalog we found 4125 have a unique mid-IR potential counterpart within 3\arcsec.3. However, only for 255 out of 4125 sources their mid-IR counterparts were detected by \wse\ in all three of its short-wavelength bands, and so we only do a comparison with this subset.

We assume that mid-IR correspondences found for the SDSSDR7WD are the real counterparts of the WDs. However in the worst scenario if none of the WDs listed in the SDSSDR7WD has mid-IR emission this will automatically confirm that their mid-IR spectral behavior is different from that of \fer\ BZBs known to date. We also computed the chance probability of having spurious associations within 3\arcsec.3 when crossmatching the All\wse\ source catalog with the SDSSDR7WD; the result is less than 4\% \citep[see e.g.,][for more details on the method adopted for the pchance calculation]{ugs1}.

To verify the reliability of our choice for the WD sample we also used { the latest release of the McCook \& Sion White Dwarf Catalog\footnote{http://www.astronomy.villanova.edu/WDCatalog/index.html} available to date} \citep[see][for details]{mccook99}. This catalog has been also used to search for dusty disks around WDs \citep{hoard13} using \wse\ observations. However in this sample the number of sources detected at 3.4$\mu$m, 4.6$\mu$m and 12$\mu$m for which is possible to built the mid-IR colors is only 91. The chance probability of spurious associations between the McCook and Sion WD catalog and the All\wse\ survey is lower than 1\%.

\subsection{Weak emission line quasars}
In the last two decades an intriguing and also elusive subclass of radio-quiet quasars was discovered: the weak emission line quasars (WELQs), featuring exceptionally weak or completely missing broad emission lines in the ultraviolet (UV) rest-frame energy range \citep[see e.g,][]{collinge05,shemmer06,ds9}. 

The original criterion to define WELQs with respect to other members of the same class (i.e., normal quasars) is based on the $EW_r$ distribution of the Ly$\alpha$ $\lambda$1216+N V $\lambda$1240 blend for a sample of quasars in the Sloan Digital Sky Survey \citep[SDSS;][]{} at redshift $z>$3, being lower than 15$\AA$ as occurs for the $>3\sigma$ weak tail of the $EW_r[Ly\alpha+NV]$ distribution. Using this definition, a first sample of $\sim$70 WELQs was defined \citep{ds9}. Since at low $z$ it is not always possible to apply the original criterion on the $EW_r[Ly\alpha+NV]$ because these emission lines lie outside the optical energy range covered by the SDSS spectrograph an equivalent criterion was defined selecting WELQs on the basis of the $EW$ of Mg II $\lambda$2800, C III] $\lambda$1909, and/ or C IV $\lambda$1549 \citep[see e.g.,][]{m14}.

When WELQs were discovered the possibility of being classified BZBs was explored and discarded \citep[see e.g.][for a recent review]{plotkin15}, however a comparison of their mid-IR properties with those of confirmed BZBs has not yet been completely performed. Here we selected a sample of WELQs to show that their mid-IR colors are generally similar to those of normal quasars but not to BZBs. Additional analyses based on the SEDs and/or broadband spectral indices (i.e., radio-to-optical $\alpha_{ro}$ and optical-to-Xray $\alpha_{ox}$) were also used to distinguish between WELQs and BL Lacs \citep{plotkin10,lane11,wu12} when multifrequency observations are available. Thus in the following sections we also investigate the broad band SEDs of RWBL candidates selected on the basis of the combination of the mid-IR colors and the optical spectra.

We considered two catalogs of WELQs, spectroscopically selected from the SDSS. The first catalog lists 73 WELQs in the redshift range between 3.03 and 5.09 \citep[][hereinafter DS9 subsample]{ds9} while the second lists 46 sources with 0.614$<z<$3.351 \citep[][hereinafter M14 subsample]{m14}. We excluded 6 sources out of these 119 WELQs, namely: SDSS J074451.36+292006.0 and SDSS J092145.37+233548.1 from the M14 subsample and SDSS J141318.86+450522.9 from DS9 subsample because they are all listed in the \bzcat\ and SDSS J104831.29+211552.2, SDSS J115326.70+361726.3, SDSS J144204.03+132916.0 again from the M14 subsample since they show a flat radio spectrum that is widely interpreted as jet signature so suggesting that their quasi-featureless optical continuum marks non-thermal synchrotron emission.

Only 90 out of the remaining 113 sources arising from the combination of the DS9 and the M14 subsamples have a counterpart in the All\wse\ catalog. These 90 sources thus constitute our WELQ reference sample to carry out the comparison with BZB mid-IR colors. We estimated the chance probability of spurious associations between the SDSSDR7WD catalog and the \wse\ survey as being lower than 1\%.

\section{BZBs, WDs and WELQs at mid-IR wavelengths}
\label{sec:analysis}
\begin{figure}[!h]
\includegraphics[height=8.cm,width=8.5cm,angle=0]{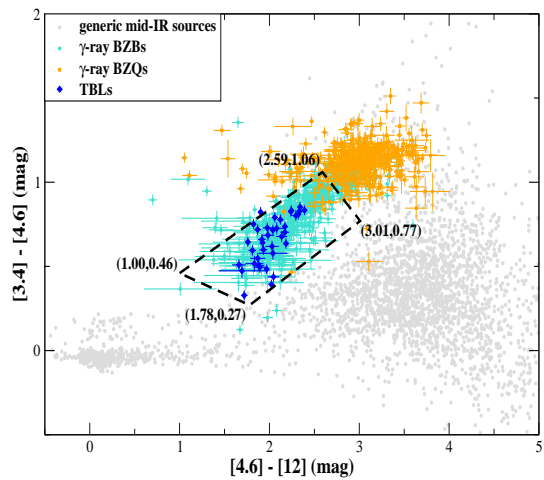}
\caption{The mid-IR color-color plot built with the \wse\ magnitudes at 3.4$\mu$m, 4.6$\mu$m and 12$\mu$m. BZBs and BZQs are displayed as cyan circles and orange squares respectively, thus corresponding to the ``\wse\ Gamma-ray Strip'' \citep{paper1}. Generic mid-IR sources (grey circles) of about 3000 sources selected at high Galactic latitudes are also show. The separation between the mid-IR colors of the \fer\ blazars and that of other sources randomly chosen appears evident. The presence of BZBs in the region of the ``\wse\ Gamma-ray Strip'' occupied by the BZQs is only limited to a handful of sources \citep[see e.g.,][for more details]{ugs1,wibrals}. The dashed polygon marks the region used to select potential RWBLs.
}
\label{fig:strip}
\end{figure}

In Fig.~\ref{fig:strip} we show the mid-IR colors built with the in the 3.4$\mu$m, 4.6$\mu$m and 12$\mu$m magnitudes of the BZBs and the BZQs in our \fer\ blazar sample in comparison with generic infrared sources selected at high Galactic latitudes ($|b|>$50\degr), to highlight their peculiar spectral behavior. Since we know that BZQs generally lie in the region occupied by ``normal'' quasars, at the top end of the so called ``\wse\ Gamma-ray Strip'' \citep{paper1}, to highlight the mid-IR color region where BZBs reside, we will only consider RWBL candidates those \wse\ sources lying in the dashed polygon shown in Fig.~\ref{fig:strip} where the contamination of the BZB region by BZQ is lower than 5\%, lacking a radio counterpart. It is worth noting that the selected region also includes TeV BL Lacs.

\begin{figure*}
\includegraphics[height=8.cm,width=8.8cm,angle=0]{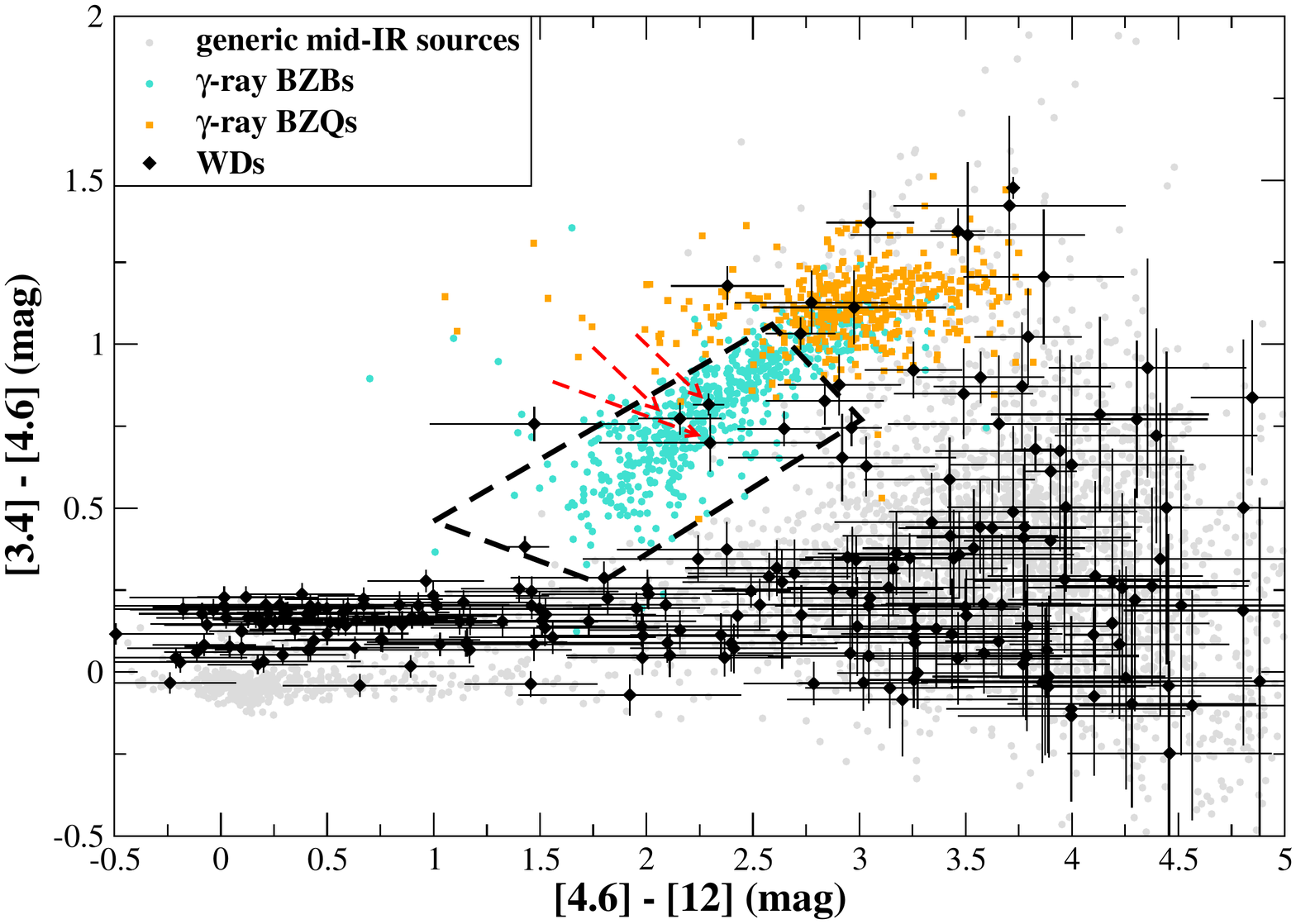}
\includegraphics[height=8.cm,width=8.8cm,angle=0]{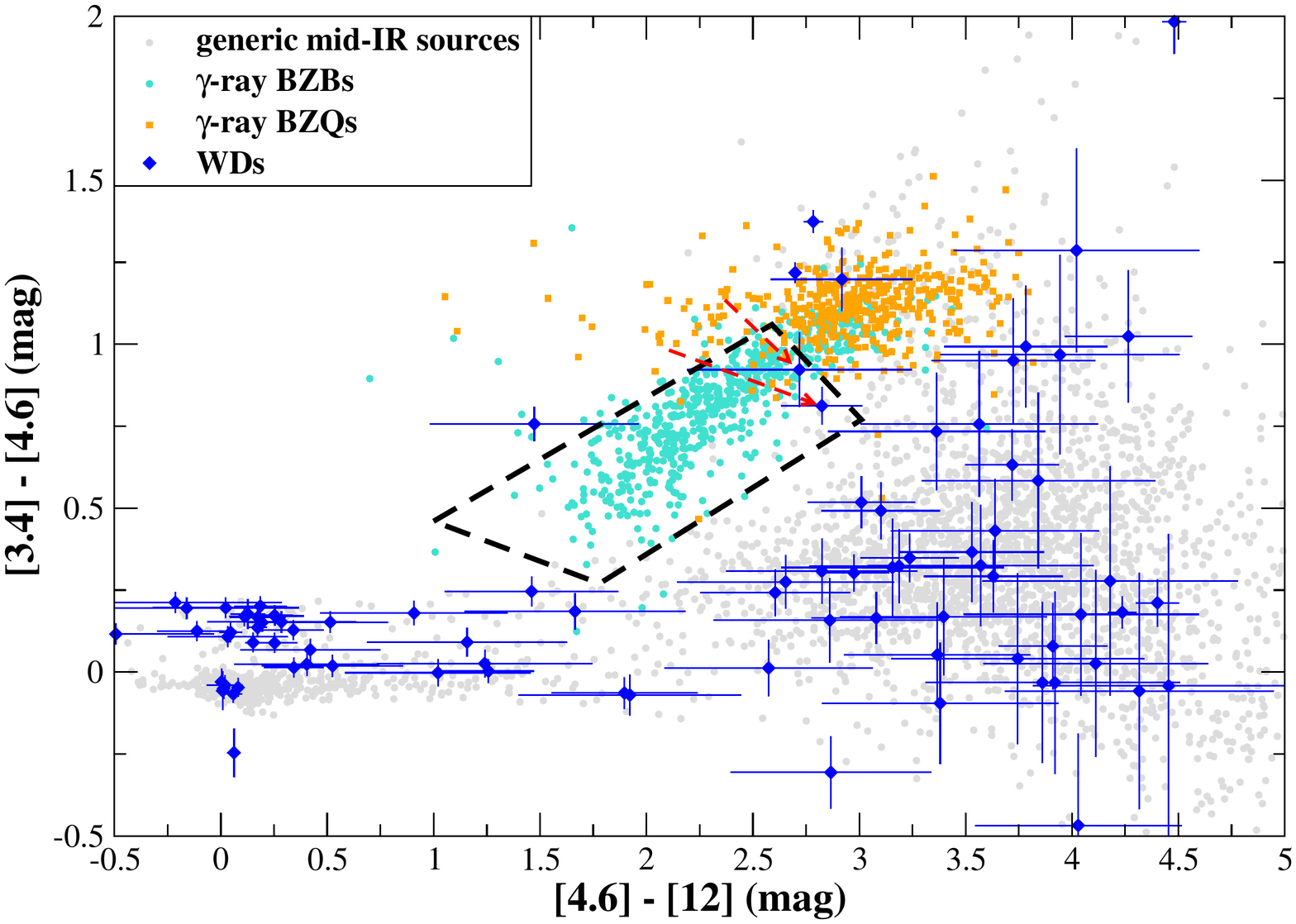}
\caption{
The comparison between the mid-IR colors of the BZBs and the BZQs as shown in Fig.~\ref{fig:strip} and those of the WDs (black and blue circles) from the SDSSDR7WD (left panel) and the McCook \& Sion WD (right panel) catalogs, respectively. In both comparisons the mid-IR colors of blazars appear well distinct from those of WDs, with the only few exceptions discussed in \S~\ref{sec:analysis} { and marked by the red arrows}. Generic mid-IR sources are also shown in the background in both panels. Uncertainties on the mid-IR colors of BZBs and BZQs are not reported here as in Fig.~\ref{fig:strip} for sake of simplicity since they are shown here for comparison only.
}
\label{fig:wds}
\end{figure*}

Fig.~\ref{fig:wds} shows the comparison between \wse\ sources associated with WDs and our blazar sample. It is evident that only 4 out of the 255 sources in the SDSSDR7WD catalog have mid-IR colors consistent with the main BZB region of the ``\wse\ Gamma-ray Strip'', while the remaining $\sim$99\% of the WDs lie well separated from this region. Three out of four sources are extragalactic, while WISE J122859.87+104032.8 has a clear WD optical spectrum, making it the only WD contaminant of the BZB region in the mid-IR color-color diagram. Moreover WISE J122859.87+104032.8 has a negative value of the $u-r$ a situation that never occurs for the known BZBs \citep{massaro12c,massaro14b}. WISE J172633.50+530300.3 also shows a featureless optical spectrum with a peculiar continuum shape; this could be due to the presence of a nearby source clearly visible in the SDSS image but not in the \wse\ one. This implies a relatively small contamination of the mid-IR color selection.

Both WISE J001736.91+145101.9 and WISE J121348.83+642520.0 are BL Lacs with a radio counterpart in addition to WISE J121348.83+642520.0 listed in the latest release of the \bzcat\ catalog. 

{ To complete our comparison between WDs and BZBs we also show the mid-IR color-color plot with the 91 sources selected out of the McCook \& Sion WD catalog in Fig.~\ref{fig:wds}. There are also in this case only two objects that could be confused as BZBs on the basis of their mid-IR colors, namely WISE J085506.13+063904.1 and WISE J115642.21+130603.9, both lying in the footprint of the SDSS. Thus in both cases we checked the optical images using the SDSS Navigate tool\footnote{http://skyserver.sdss.org/dr12/en/tools/chart/navi.aspx}. We found that about 6\arcsec\.6 from the location of WISE J085506.13+063904.1 there is a clear WD, namely SDSS J085506.62+063904.7, for which an optical spectrum is available being most likely the star listed in the McCook \& Sion WD catalog and lacking a \wse\ counterpart. This mis-association is due to the positional uncertainty of the sources in the McCook \& Sion WD catalog. Thus we conclude that WISE J085506.13+063904.1 is a simple contaminant unrelated to the WD selection. Then for the latter, WISE J115642.21+130603.9 the lack of its optical spectrum combined with the uncertainties of the SDSS photometric observations does not allow us to verify its nature. However, this object is classified as {\it galaxy} thus unlikely to be a WD that can contaminate our RWBL selection.}

The comparison between mid-IR colors of BZBs and WDs prove the usefulness of the mid-IR color-diagram to distinguish the two classes while searching for RWBLs, once optical spectra are available.

\begin{figure}
\includegraphics[height=8.cm,width=8.8cm,angle=0]{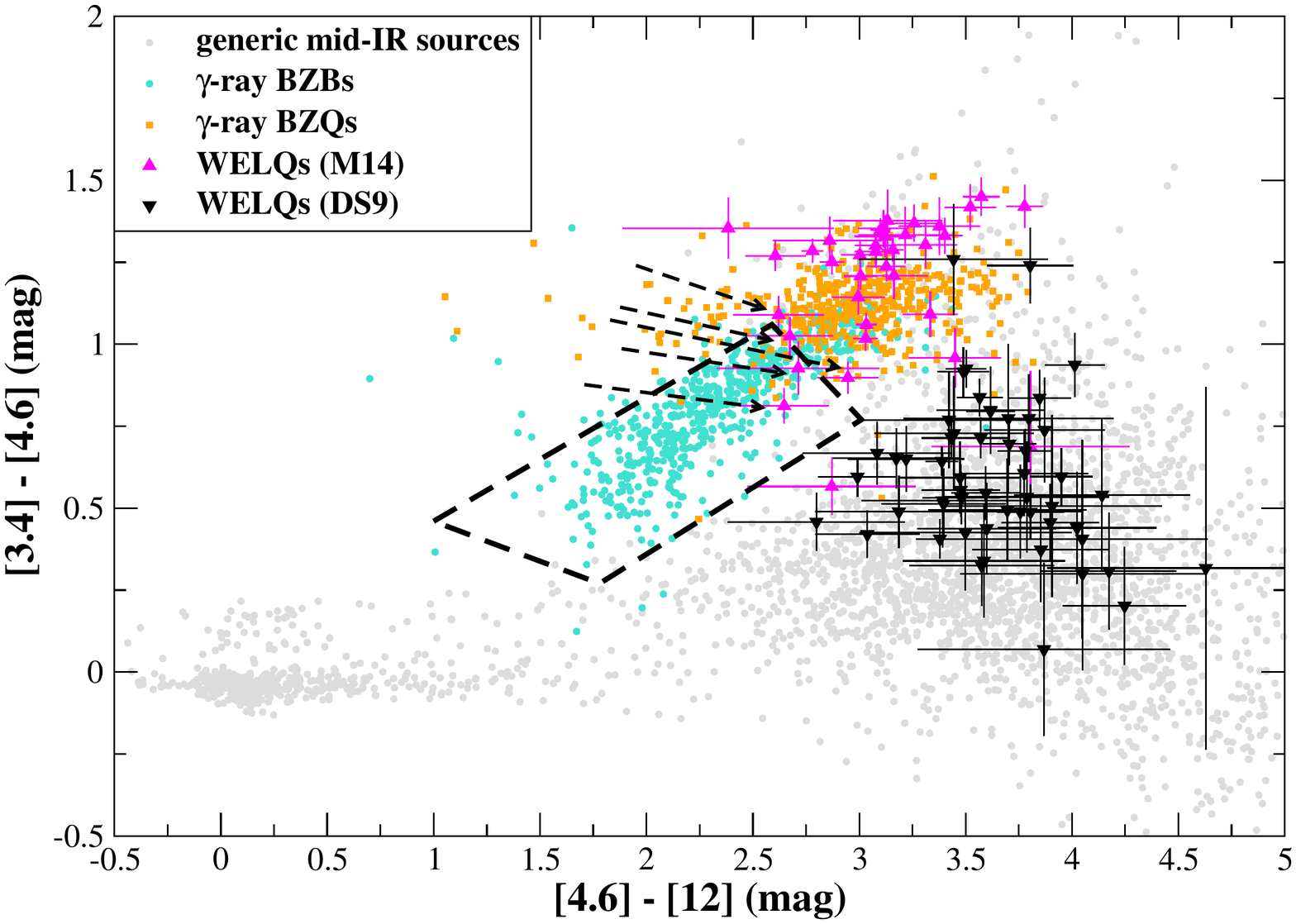}
\caption{
The comparison between the mid-IR colors of the BZBs and the BZQs as shown in Fig.~\ref{fig:strip} and those of the WELQs. The WELQs belonging to the M14 subsample are shown in magenta triangles (up) while those of the DS9 subsample in black (down). In both comparisons the mid-IR colors of blazars appear well distinct from those of WELQs, with the only few exceptions discussed in \S~\ref{sec:analysis} { and marked here by the black arrows}. Generic mid-IR sources are also shown in the background in both panels. Uncertainties on the mid-IR colors of BZBs and BZQs are not reported here as in Fig.~\ref{fig:strip} for sake of simplicity since they are shown here for comparison only.
}
\label{fig:wlqs}
\end{figure}

Fig.~\ref{fig:wlqs} shows the mid-IR comparison between the WELQs and the BZBs. Here too is clear the neat separation between these two classes. It is also worth noting that sources in the DS9 subsample appear to lie in a different region of the color parameter space with respect to the M14 subsample. The reason underlying this effect is due to their different redshift distributions. We note that there is also a marginal difference between the mid-IR colors of the WELQs in the M14 sample and that of the BZQs but it is less useful to search for RWBLs hidden among this sample.

In this case we highlight the fact that two WELQs lie below the threshold marking the BZB region of the color-color plot and additional three objects are close to it. These two sources are WISE J001514.87-103043.4 and WISE J150427.69+543902.2; both showing MgII$\lambda$2798 emission line of rest-frame equivalent width $EW_r$ consistent with the 5\AA\ threshold, usually adopted to classify BL Lacs, within 3$\sigma$, thus being ``border line cases''. The remaining three sources, belonging to the M14 subsample (see Fig.~\ref{fig:rwbls}, where they are shown in magenta), all have a radio counterpart at 1.4 GHz in the Faint Images of the Radio Sky at Twenty centimeter \citep[FIRST;][]{helfand15} catalog, but the lack of additional radio data did not allow us to verify the flatness of their radio spectra, although they are definitively not RWBLs. The location of all five sources with respect to that of the BZBs in the mid-IR color-color plot is shown in Fig.~\ref{fig:rwbls}.

Finally, we conclude that both WDs and WELQs have distinct mid-IR colors with respect to the BL Lac population and [3.4]-[4.6]-[12]$\mu$m color-color diagram can be used as diagnostic tool to confirm the source nature having featureless optical spectra. Sources that appear to be contaminants of the mid-IR selection procedure are ruled out when \wse\ data are combined with optical spectroscopic observations.

\section{Radio Weak BL Lacs}
\label{sec:rwbls}
In the literature there are five sources tentative classified as RWBLs. 

\begin{enumerate}
\item 2QZ J215454.3-305654 \citep{londish02} for which several multifrequency follow up observations have been performed to date was the first. However its nature is still uncertain since it could be a RWBL or a ``lineless'' quasar \citep{londish04}. 

\item Paggi et al. (2014), carried out optical spectroscopic observations of blazar candidates, potential UGS counterparts and discovered WISE J064459.38+603131. This mid-IR sources lies within the positional uncertainty region of 2FGL J0644.6+6034.7 (3FGL J0644.6+6035). This object shows a single optical emission line of EW consistent with the 5\AA\ threshold, arbitrarily set in the literature for the BL Lac class definition, similar to the two cases discussed at the end of \S~\ref{sec:analysis}. 

\item In 2015 WISE J173052.85-035247.2 was discovered during the optical spectroscopic follow up observations of UGSs \citep{ricci15}. This source could be the potential counterpart of 2FGL J1730.6-0353 (a.k.a.  3FGL J1730.6-0357). 

\item More recently Landoni et al. (2016) claimed that also SDSS J004054.65-0915268 could be a tentative RWBL. 

\item Finally, Marchesini et al. (2016) discovered the last known candidate of RWBL, WISE J141046.00+740511.2. This source was also found while carrying out spectroscopic observations of UGS listed in the First Fermi-LAT Catalog of Sources above 10 GeV \citep[1FHL;][]{ackermann13} selected on the basis of the \swf\ X-ray observations \citep{landi15}. It has been indicated as potential counterpart of 1FHL J1410.4+7408 (a.k.a. 3FGL J1410.9+7406).
\end{enumerate}
\begin{figure*}[]
\includegraphics[height=8.cm,width=8.8cm,angle=0]{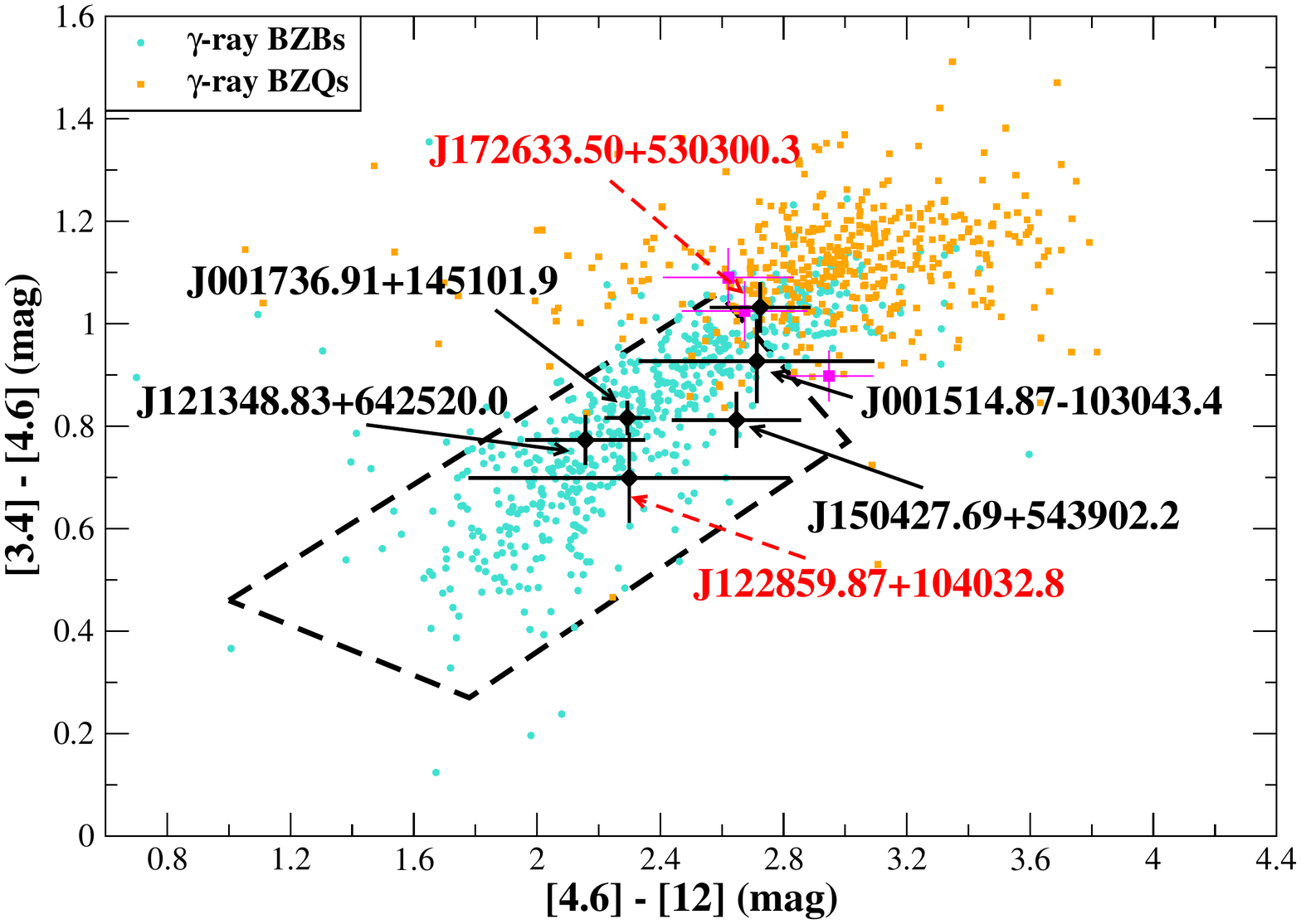}
\includegraphics[height=8.cm,width=8.8cm,angle=0]{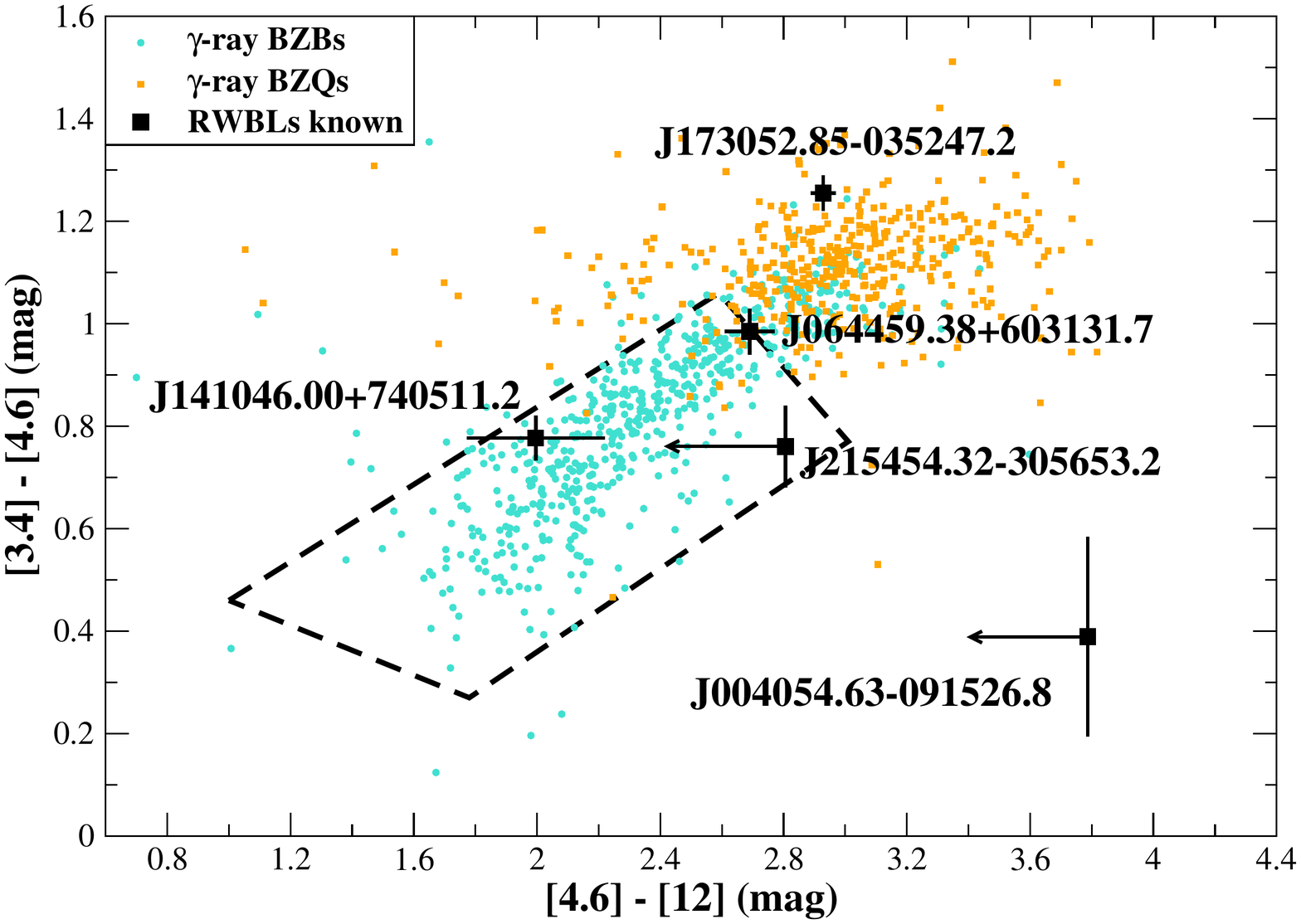}
\caption{
Left panel) The mid-IR color-color plot where the positions of the blazars recognized in the WD and the WELQ samples are highlighted with respect to the \fer\ blazars. WISE J122859.87+104032.8 and WISE J172633.50+530300.3 are marked with a red arrow since additional optical information allow us to discard them as potential RWBLs. The former has the unusual negative value of the $u-r$ never seen for BL Lacs while the mid-IR colors of the latter are contaminated by the presence of a nearby source visible in the optical image (see \ref{sec:analysis} for more details). Right panel) The positions of the candidates RWBLs in the mid-IR color-color plot are shown in comparison to that of \fer\ blazars belonging to the ``\wse\ Gamma-ray Strip'' and those of generic mid-IR sources (same color code of previous figures). Source details for the objects indicated in both panels are given in \S~\ref{sec:analysis}. Uncertainties on the mid-IR colors of BZBs and BZQs are not reported here as in Fig.~\ref{fig:strip} for sake of simplicity since they are shown here for comparison only.
}
\label{fig:rwbls}
\end{figure*}

In Fig.~\ref{fig:rwbls} we show the positions of the five RWBLs overlay to the ``\wse\ Gamma-ray Strip'' subregion of BZBs.
SDSS J004054.65-0915268 (a.k.a WISE J004054.63-091526.8) has an upper limit at 12$\mu$m and being a high $z$ source it seems to be more consistent with the location of the DS9 subsample of WELQs rather than the BZB population. A similar situation occurs for 2QZ J215454.3-305654 (a.k.a. J215454.32-305653.2) that only a follow up observations at 12$\mu$m could potentially reveal if the source is consistent with the mid-IR colors of the BZBs. WISE J064459.38+603131.7, being selected on the basis of its mid-IR colors is definitively consistent with the BZB population on the 3.4$\mu$m - 4.6$\mu$m - 12$\mu$m diagram, while WISE J173052.85-035247.2 appears to have such colors more similar to the WELQs listed in the M14 subsample. 

The remaining case of WISE J141046.00+740511.2 is the most intriguing one. This object is potentially associated to a 1FHL object and its potential \wse\ counterpart lies in the region of the ``\wse\ Gamma-ray Strip'' where all known TeV BL Lacs are located \citep{massaro13d,arsioli15}. 
The lack of any radio counterpart in both the NRAO VLA Sky Survey \citep[NVSS;][]{condon98} and the FIRST catalog \citep{helfand15} that cover the field of WISE J141046.00+740511.2 makes this source the most promising candidate to be the first, genuine, RWBL. Nevertheless is it worth emphasizing that the distance between this mid-IR source and the gamma-ray position of 3FGL J1410.9+7406 is $\sim$90\arcsec well below the average of the associated \fer\ blazars listed in the \bzcat\ corresponding to $\sim$130\arcsec. At the angular separation of $\sim$90\arcsec, the association probabilities of counterparts listed in the latest releases of the \fer\ catalogs is typically grater than 99.5\% \citep{ackermann15}.

Applying this mid-IR color-based analysis, in combination with optical spectra, we conclude that WISE J064459.38+603131 and WISE J141046.00+740511.2, lying within the positional uncertainty regions of two distinct UGSs listed in the 3FGL, are the first two RWBLs detected so far. Both of them can also be associable with 1FHL sources thus strengthening our interpretation, since the high energy sky above 10 GeV is mostly dominated by BZBs. In the following we will also show that additional multifrequency observations can strengthen our results.

\section{mid-IR variability}
\label{sec:variability}
BL Lac objects are known to be among the most variable extragalactic sources at all frequencies, showing variations either in spectral shape or in flux/luminosities. This peculiar behavior could in principle mine the mid-IR color selection proposed here and in our previous works. However this situation does not seem to occur. 

\begin{figure}[!ht]
\includegraphics[height=8.cm,width=8.8cm,angle=0]{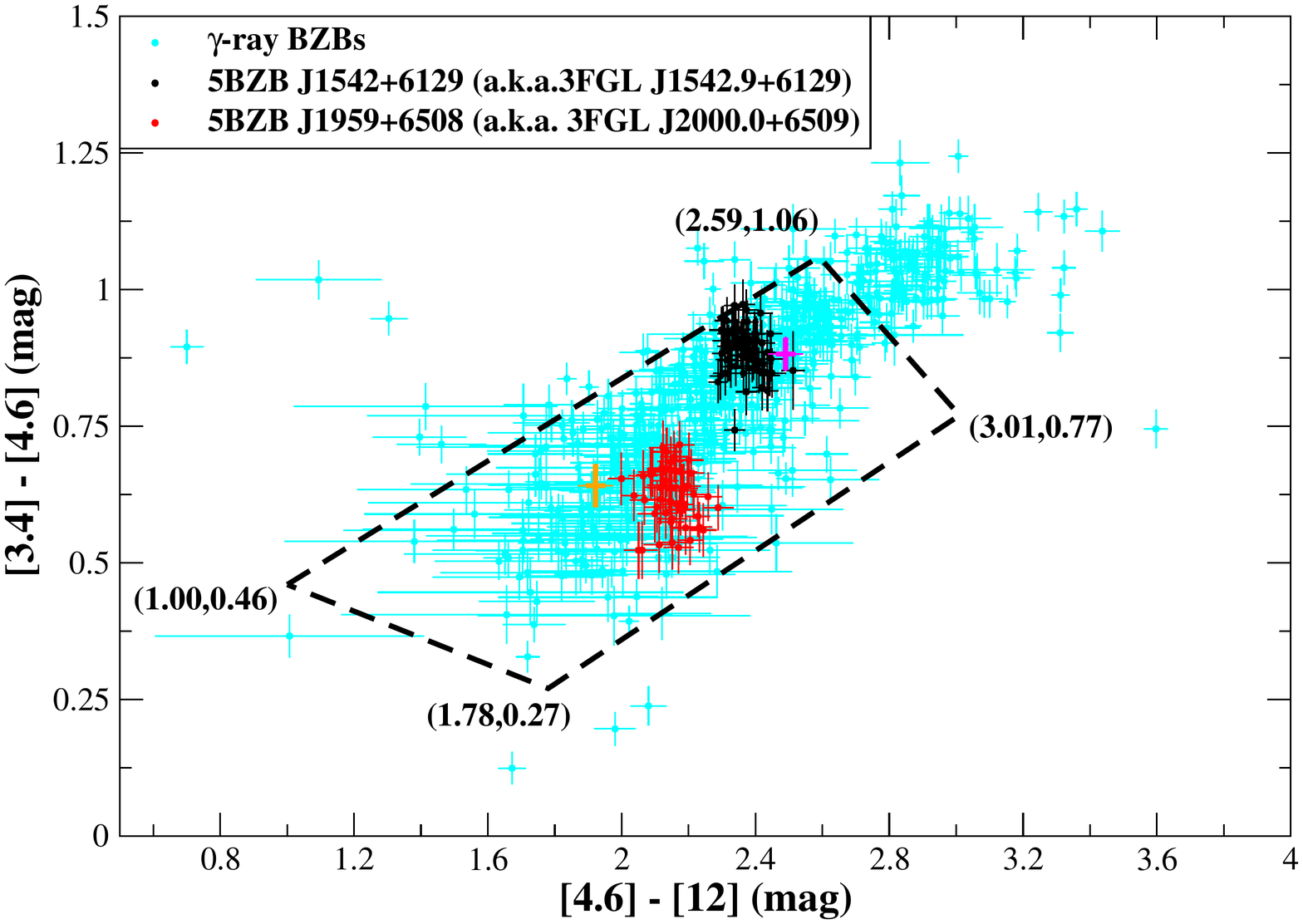}
\caption{The mid-IR color-color plot built with the \wse\ magnitudes at 3.4$\mu$m, 4.6$\mu$m and 12$\mu$m. BZBs are displayed as cyan circles and correspond to the ``\wse\ Gamma-ray Strip'' \citep{paper1}. We show two cluster of points relative to the \wse\ single epoch observations of the BZBs: 5BZBJ1542+6129 (black circles) and 5BZBJ1959+6508 (red circles). It appears evident that even mid-IR color variation states of these two BL Lac objects are always consistent with the boundaries of the ``\wse\ Gamma-ray Strip''. The two crosses point to the location of 5BZBJ1542+6129 (magenta) and 5BZBJ1959+6508 (orange) on the mid-IR color diagram when integrated values for the entire \wse\ all-sky survey are used.
} 
\label{fig:moving1}
\end{figure}

BL Lacs appear to be variable in the mid-IR band \citep[see e.g.,][]{paper2} although their color variations still reside within the boundaries of the ``\wse\ Gamma-ray Strip'' \citep[see][for more details]{buson16}. Here we show an example of two BZBs: 5BZBJ1542+6129 and 5BZBJ1959+6508, the latest also detected at TeV energies \citep{aharonian03}, for which more than 50 \wse\ observations are available, when both sources are detected at least in the first three mid-IR filters at 3.4$\mu$m, 4.6$\mu$m and 12$\mu$m. In Fig.~\ref{fig:moving1} it is clear how both BZBs during their single-epoch \wse\ observations move in the mid-IR color-color plot in the same region occupied by the BL Lac objects. In addition their colors are also consistent with the boundaries chosen for our RWBL selection procedure.

\begin{figure*}
\includegraphics[height=8.cm,width=8.8cm,angle=0]{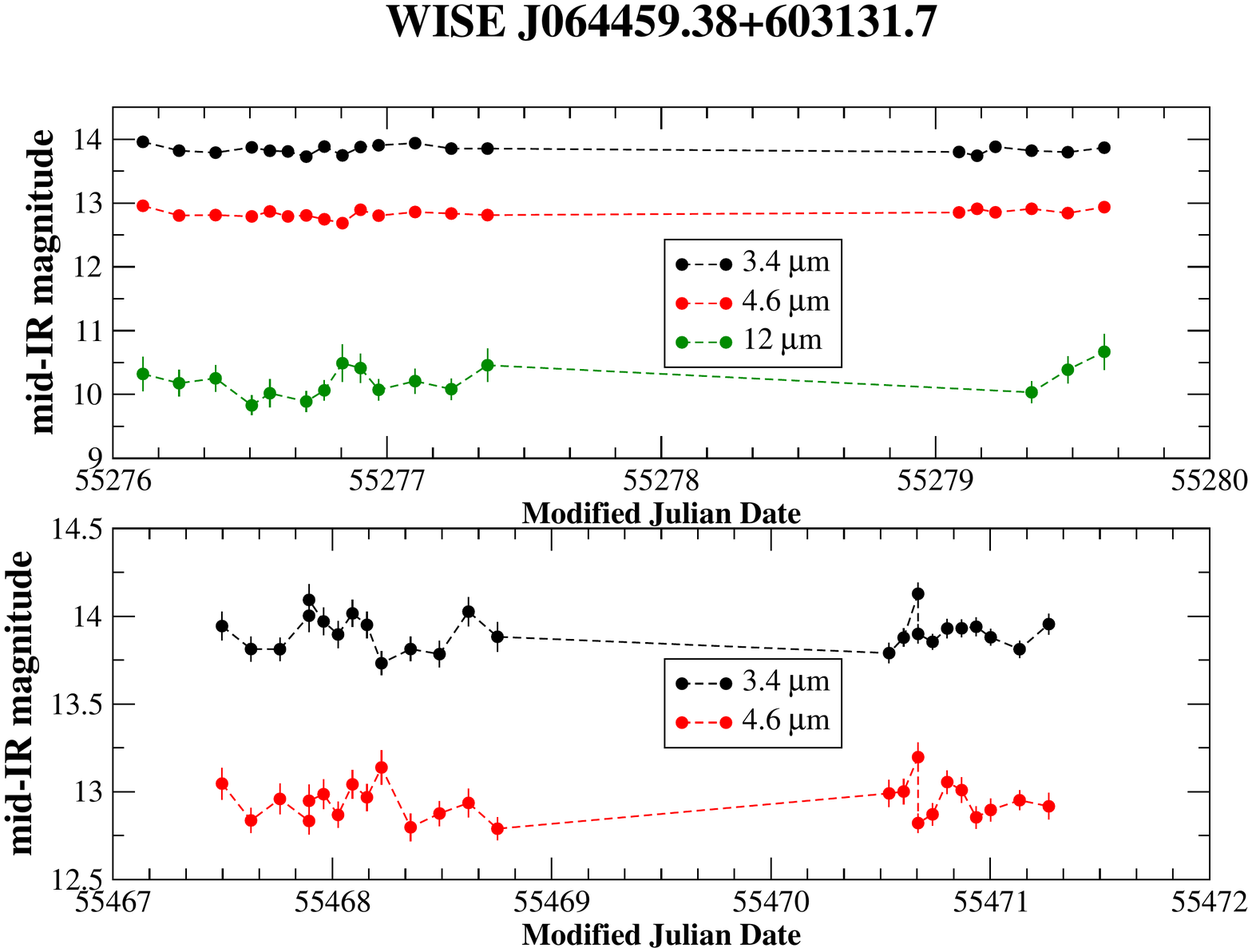}
\includegraphics[height=8.cm,width=8.8cm,angle=0]{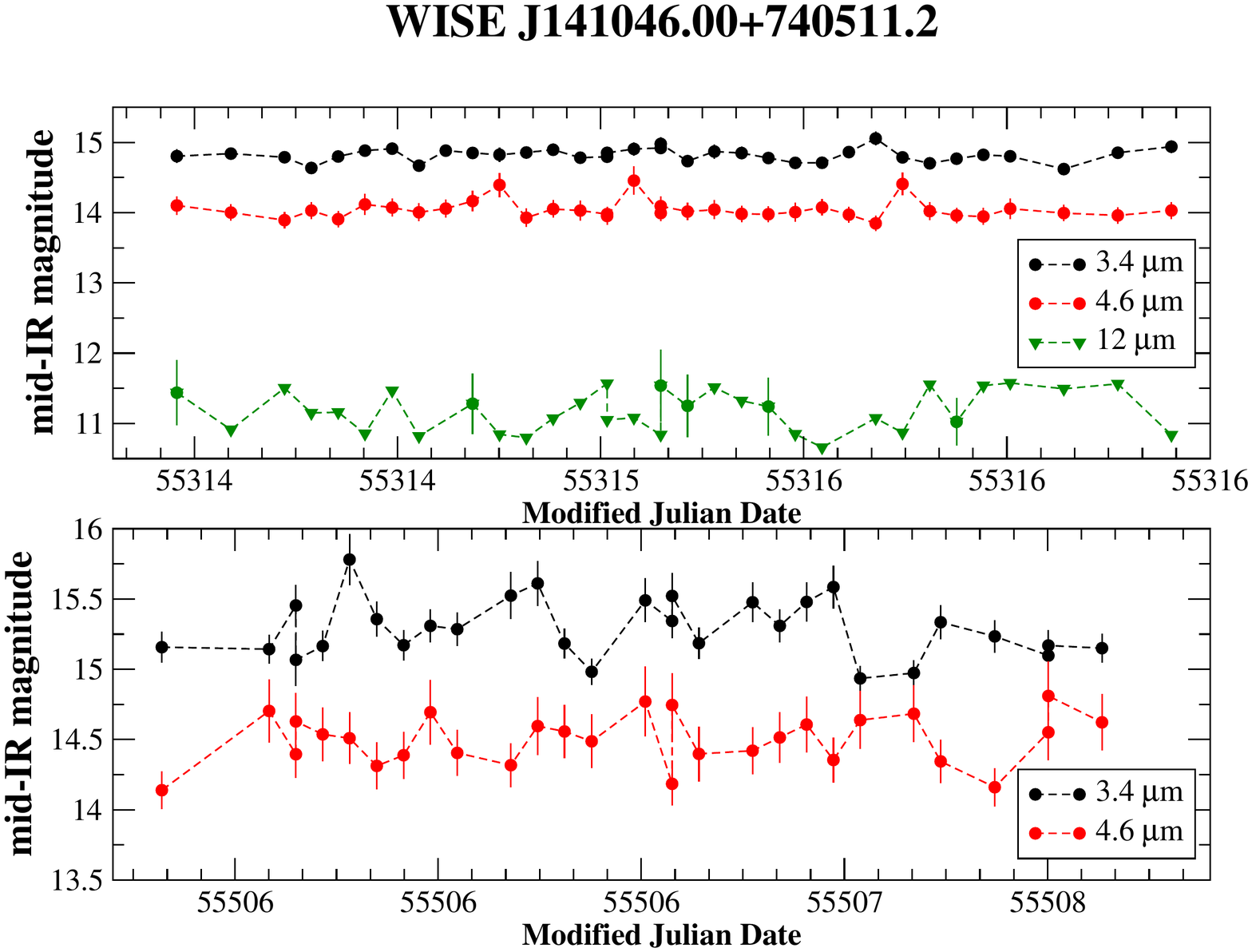}
\caption{The light curve of WISE J064459.38+603131 (left panel) and WISE J141046.00+740511.2 (right panel), candidate RWBLs, tentatively associated to 3FGL and 1FHL \fer\ sources. These are \wse\ single-epoch observations for two different time periods. Mid-IR flux density is expressed using the \wse\ magnitude (not corrected for absorption), as function of time in Modified Julian Date. Each epoch (top and lower panel) correspond to approximately 2 days. Infrared variability is clearly notable for both sources. Triangles are used to mark \wse\ observations that corresponds to upper limits at a given wavelength.
}
\label{fig:lightcurves}
\end{figure*}

\begin{figure*}
\includegraphics[height=8.cm,width=8.8cm,angle=0]{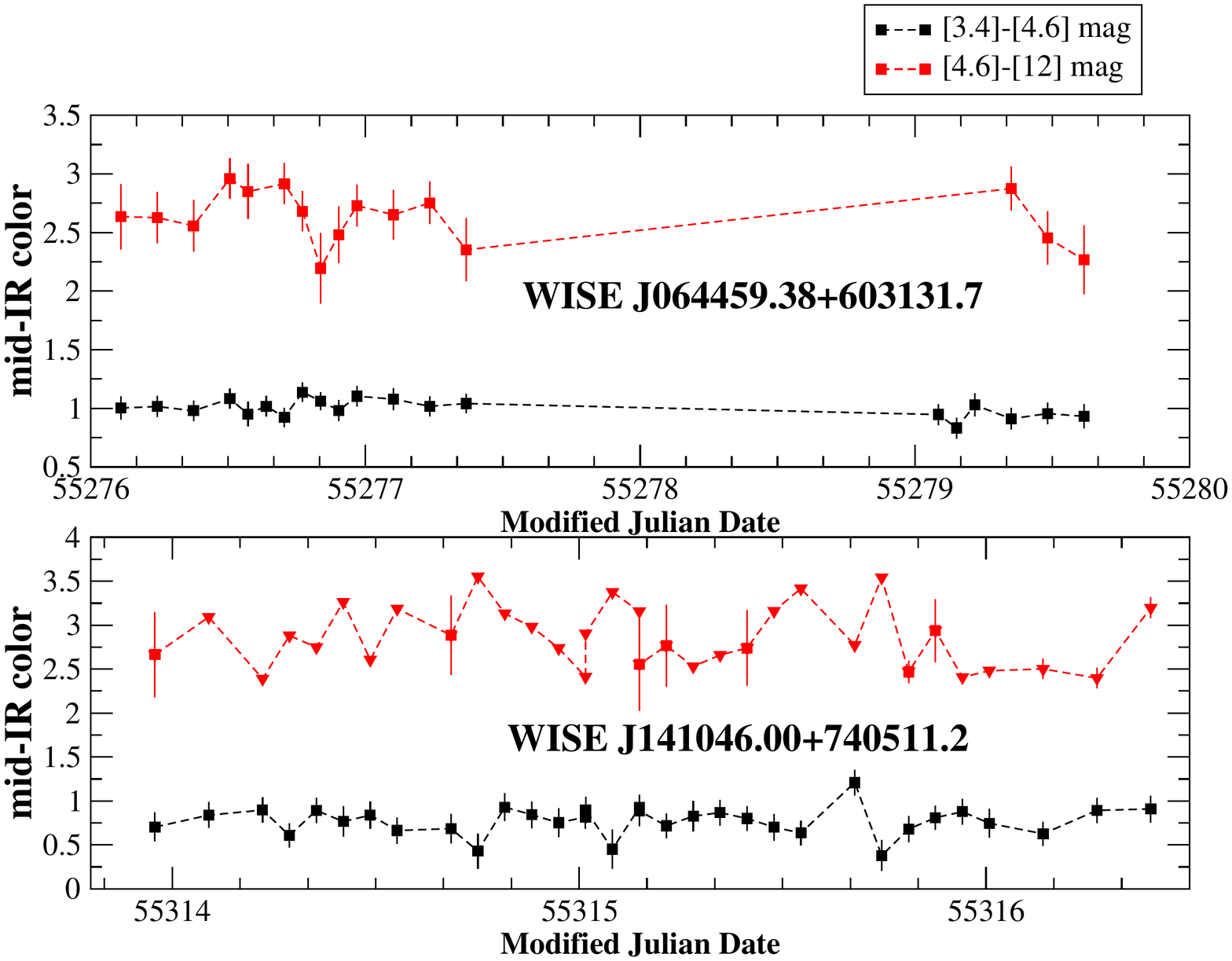}
\includegraphics[height=8.cm,width=8.8cm,angle=0]{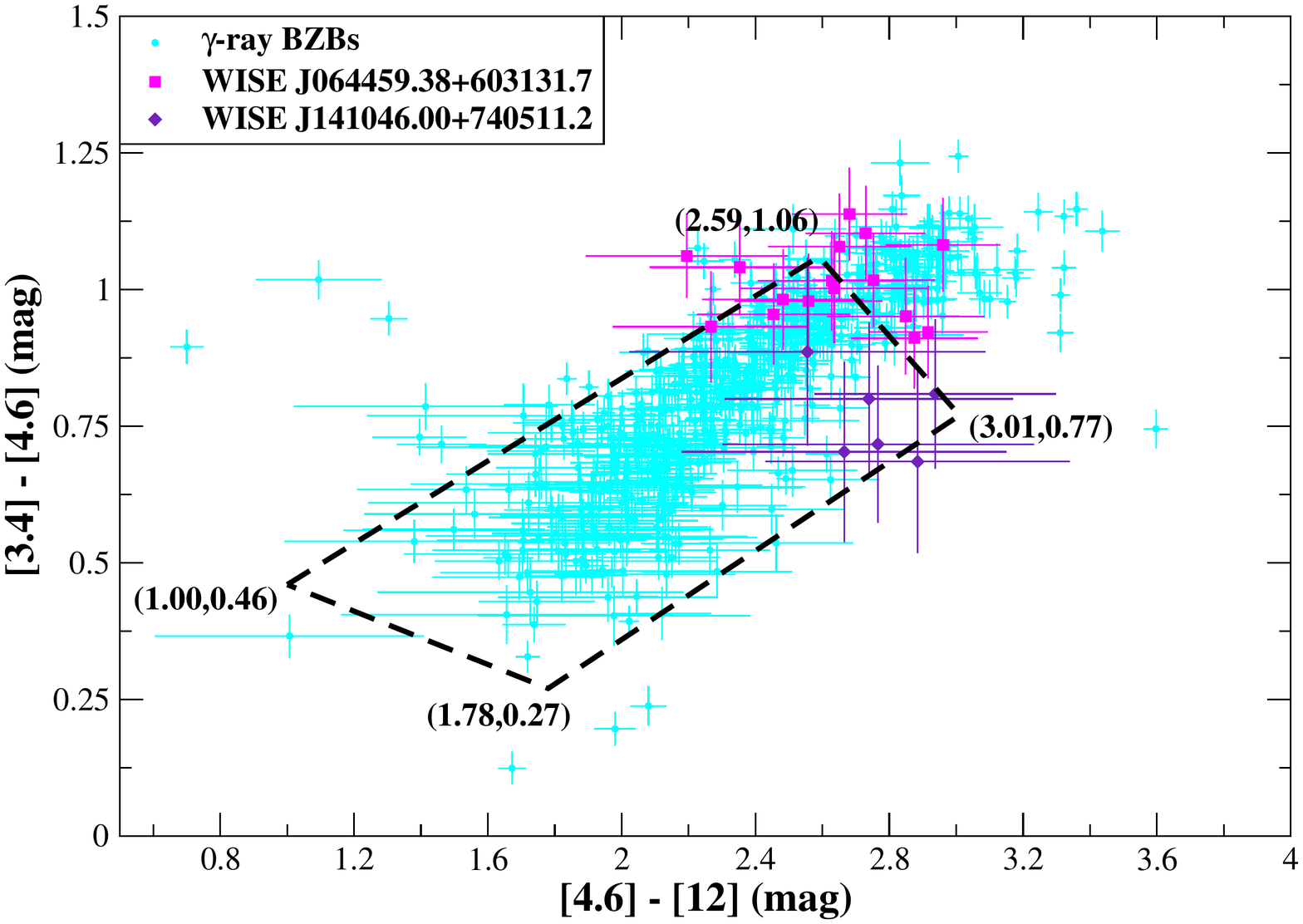}
\caption{Left panel) Mid-IR colors reported as function of time in Modified Julian Date for both RWBL selected in our analysis: WISE J064459.38+603131 (top) and WISE J141046.00+740511.2 (bottom). For WISE J064459.38+603131 we only considered \wse\ observations when the source is detected in all the three filters while for WISE J141046.00+740511.2 we computed the [4.6]-[12]$\mu$m color-lightcurve also using the upper limits (red triangles down). Right panel) Same of Figure~\ref{fig:moving1} but for the two RWBL selected: WISE J064459.38+603131 (purple squares) and WISE J141046.00+740511.2 (magenta diamond). The variations found in the single-epoch \wse\ observations when both sources are detected { in the first three mid-IR filters is clearly consistent with the boundaries of the portion of the ``\wse Gamma-ray strip'' for WISE J141046.00+740511.2 and partially in agreement for WISE J064459.38+603131.}
}
\label{fig:colorcurve}
\end{figure*}

According to the \wse\ catalog one of our best RWBL candidate: the WISE J141046.00+740511.2 is flagged as variable in the mid-IR. This is in agreement with the fact that such emission it is unlikely due to dust emission, as occurs in normal quasars or in white dwarfs \citep{debes11}. Fig.~\ref{fig:lightcurves} shows the \wse\ magnitude at 3.4$\mu$m, 4.6$\mu$m and 12$\mu$m as function of time for two different epochs; mid-IR variability, not only between the two observing periods, but also in each single set of observations is found. We also present the lightcurve for the second RWBL candidate WISE J064459.38+603131 that clearly shows infrared variability being detected more times at 12$\mu$m during the \wse\ single-epoch observations with respect to WISE J141046.00+740511.2.
In Fig.~\ref{fig:colorcurve} we also show their mid-IR color variability as function of time. In the same figure we also present the single epoch observations when both sources are detected in the first three \wse\ filters overlaid to the ``\wse\ Gamma-ray Strip''. { As for the other two BZBs mentioned above these variations are clearly consistent with the region chosen to select RWBLs for WISE J141046.00+740511.2 while for WISE J064459.38+603131 there is partial agreement}. It is worth mentioning that the location of integrated mid-IR colors for both WISE J064459.38+603131 and WISE J141046.00+740511.2 could appear different from that where they exhibit variability in the mid-IR color diagram. The underlying reason is that there are many \wse\ single-epoch observations when sources (see Fig.~\ref{fig:lightcurves}) are not detected that contribute to the color change for the integrated values. It is also worth noting that for the two RWBL candidates the uncertainties on the \wse\ magnitudes are greater than for the two bright BZBs selected above as example.

On the basis of this investigation of mid-IR variability and spectral behavior we conclude that the evidence supports the interpretation of these sources as RWBLs.

\section{\swf\ observations}
\label{sec:swift}
Several analyses were carried out to study the nature of WELQs in the X-rays \citep[see e.g.,][]{shemmer09,wu12}. These observations were also used to build their broad band spectral indices and/or their SEDs \citep[see e.g.,][]{plotkin10} to verify their potential similarities with BL Lac objects.
Since the two sources identified here as RWBLs were discovered during a campaign aimed to search for UGS counterparts and all the \fer\ UGSs are in a \swf\ observational program \citep{stroh13}\footnote{http://www.swift.psu.edu/unassociated/}, we reduced the \swf\ data available to date to build the RWBL SEDs.

WISE J064459.38+603131 was observed by \swf\ both with the X-Ray Telescope (XRT), for $\sim$ 3 ksec, and with the Ultraviolet/Optical Telescope (UVOT) instruments on board but for the latter the only ultraviolet filter used was M2. On the other hand, for WISE J141046.00+740511.2, there are optical data in the U band and all the UV filters available, together with about 11 ksec of X-ray data.

Here adopted the standard XRT and UVOT data reduction procedure, extensively used in our previous analyses \citep[see e.g.][and references therein]{specevbl,BATcan,3cunid} while only the very basic details are described below. XRT data were processed with the XRTDAS software package (v.3.0.0) developed at the ASI Science Q9 Data Center (ASDC) and distributed within the HEASOFT package (v.6.18) by the NASA High Energy Astrophysics Archive Research Center (HEASARC). All the XRT observations were carried out in the photon counting (PC) readout mode. Event files were calibrated and cleaned applying standard filtering criteria with the {\it XRTPIPELINE} task and using the latest calibration files available in the Swift CALDB. Events in the energy range 0.3--10 keV with grades 0-12 were selected in the analysis according to the standard guidelines. Exposure maps were also created using {\it XRTPIPELINE}. The \swf\ observations available for both sources have variable exposures,  we merged cleaned event files obtained with the above procedure using {\it XSELECT} task and considering only \swf\ observations with telescope aim point falling in a circular region of 12 arcmin radius centered in the median of the individual aim points, to get a uniform exposure. The corresponding merged exposure maps were then generated by summing the exposure maps of the individual observations with {\it XIMAGE} package. Source detection of in the XRT images was performed using the detection algorithm {\it DETECT}. 

We collected 13 counts for WISE J064459.38+603131 in the merged XRT event file while there are 36 counts for WISE J141046.00+740511.2, making both sources well detected above the 5$\sigma$ threshold. However the poor number of counts available, corresponding to a count rate of 0.004 cts/s and 0.003 cts/s, respectively, did not permit us to carry out a detailed spectral analysis. Thus assuming an absorbed power-law model with a photon index of 2.2 and with the Galactic hydrogen column density \citep{kalberla05} and estimated the unabsorbed X-ray flux of in the 0.5-10 keV for WISE J064459.38+603131 and of for WISE J141046.00+740511.2 using WebPIMMS\footnote{https://heasarc.gsfc.nasa.gov/cgi-bin/Tools/w3pimms/w3pimms.pl}.

Details on the UVOT reduction procedure used here are described in Massaro et al. (2008b) and Maselli et al. (2008). We performed the photometric analysis using a standard task {\it UVOTSOURCE} available in the HEASOFT package (v.6.18). Counts were extracted from a 5 arcseconds radius circular region in the V, B, and U filters as well as for the UV filters (UVW1,UVM2, and UVW2). The count rate was corrected for coincidence loss and the background subtraction was performed by estimating its level in an offset annular region centered on the source position with inner radius of 7.5 arcseconds and outer radius of 15 arcseconds. Since estimate of magnitude errors is complex due to possible instrumental systematics and calibration, here we only showed a typical uncertainty of 5\% for all filters.

The values of the observed AB magnitude for WISE J064459.38+603131 in the UV band at 1.36$\times$10$^{15}$Hz is 19.69 mag (M2 UVOT filter) while for WISE J141046.00+740511.2 the AB magnitudes measured in the UVOT filters are: 19.23 mag (U), 19.85 mag (W1), 20.24 mag (M2) and 20.31 mag (W2), respectively.

\begin{figure}
\includegraphics[height=6.6cm,width=8.8cm,angle=0]{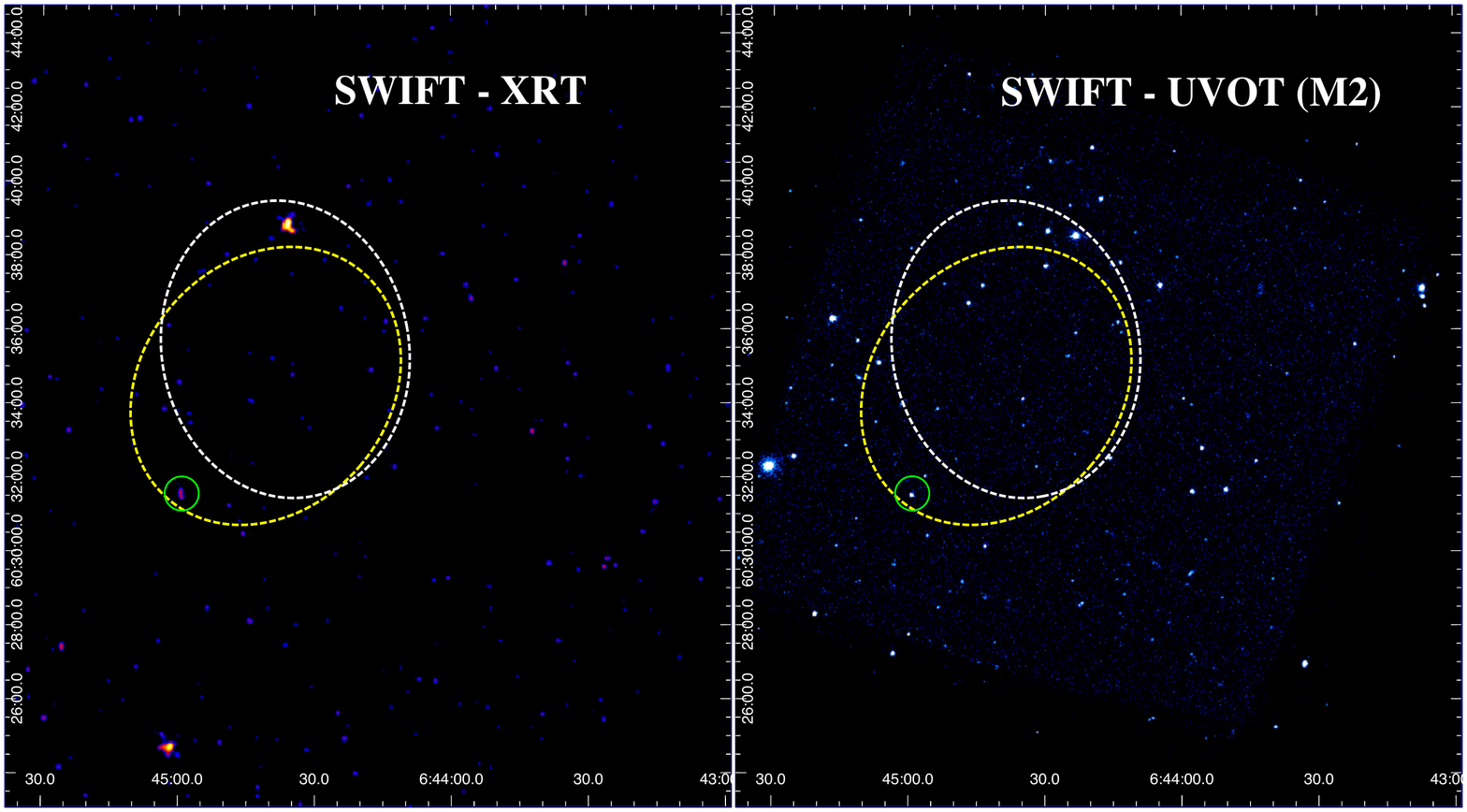}
\includegraphics[height=6.6cm,width=8.8cm,angle=0]{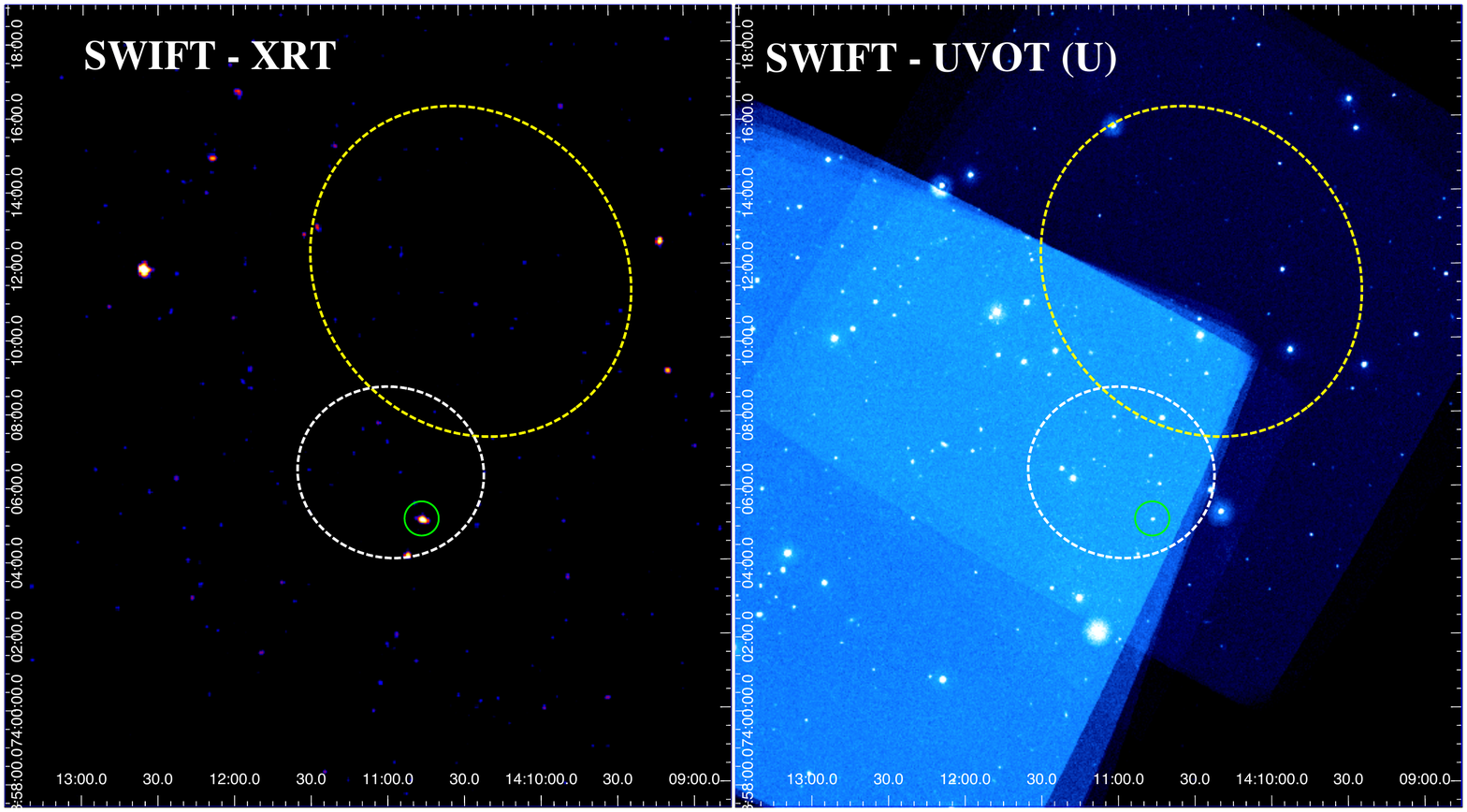}
\caption{Top panels) The XRT (left) and the UVOT (M2 filter) images of the field where WISE J064459.38+603131 lies. The yellow dashed ellipse marks the positional uncertainty region of the closest 2FGL unassociated source while the white one the error ellipse from the 3FGL catalog, both at 95\% level of confidence; { green circles mark the location of the X-ray counterpart.} Bottom panel) Same of top panels for WISE J141046.00+740511.2. The UVOT image in this case is in the optical energy range (i.e., U filter) instead of the UV.
}
\label{fig:swift}
\end{figure}

In Fig.~\ref{fig:swift} we show the XRT event file in the 0.3-10 keV energy range and one of the UVOT images for both the RWBL selected. We overlaid to the \swf\ images the positional uncertainty region, at 95\% level of confidence, from the 2FGL and the 3FGL catalogs for the nearby UGSs. WISE J064459.38+603131 lies within the error ellipse of the 2FGL, as occurred when it was selected \citep{massaro12a} but not within the one of the 3FGL, while WISE J141046.00+740511.2 that is associable to the 3FGL source, it was not found a potential counterpart of the 2FGL UGS. 

\swf\ observations, both in the UV and in the X-ray energy ranges are used in the following to build the SEDs for both the RWBLs selected.

\section{Spectral Energy Distributions}
\label{sec:seds}
For the two RWBL selected we also built the SEDs to check if they are similar or not to those of WELQs \citep{lane11}. Analysis of the SEDs is equivalent to comparing broad band spectral indices that could give an additional indication of the differences between BL Lacs and WELQs \citep[see e.g][]{wu12}.

In the case of WISE J064459.38+603131, the source has an upper limit in the NVSS catalog of 0.13 mJy/beam while WISE J141046.00+740511.2 1.69 mJy/beam. The former is detected in all the \wse\ bands and in the 2MASS. From the \swf\ observations we also got an estimate of the UV flux density in the M2 filter and the X-ray flux. The latter source still has a \wse\ counterpart detected at all mid-IR frequencies but it does not have a 2MASS correspondence. \swf\ observations provided us flux density estimates in the optical (U) and in all the UV filters (W1, M2, W2) together with the X-ray one.

We corrected the all the infrared, optical and ultraviolet flux densities for the Galactic absorption following the Draine (2003) and the reddening corrections of Schlafly \& Finkbeiner (2011), with the only exception for the \wse\ magnitudes at 12$\mu$m and 22$\mu$m.

\begin{figure*}
\includegraphics[height=6.6cm,width=8.8cm,angle=0]{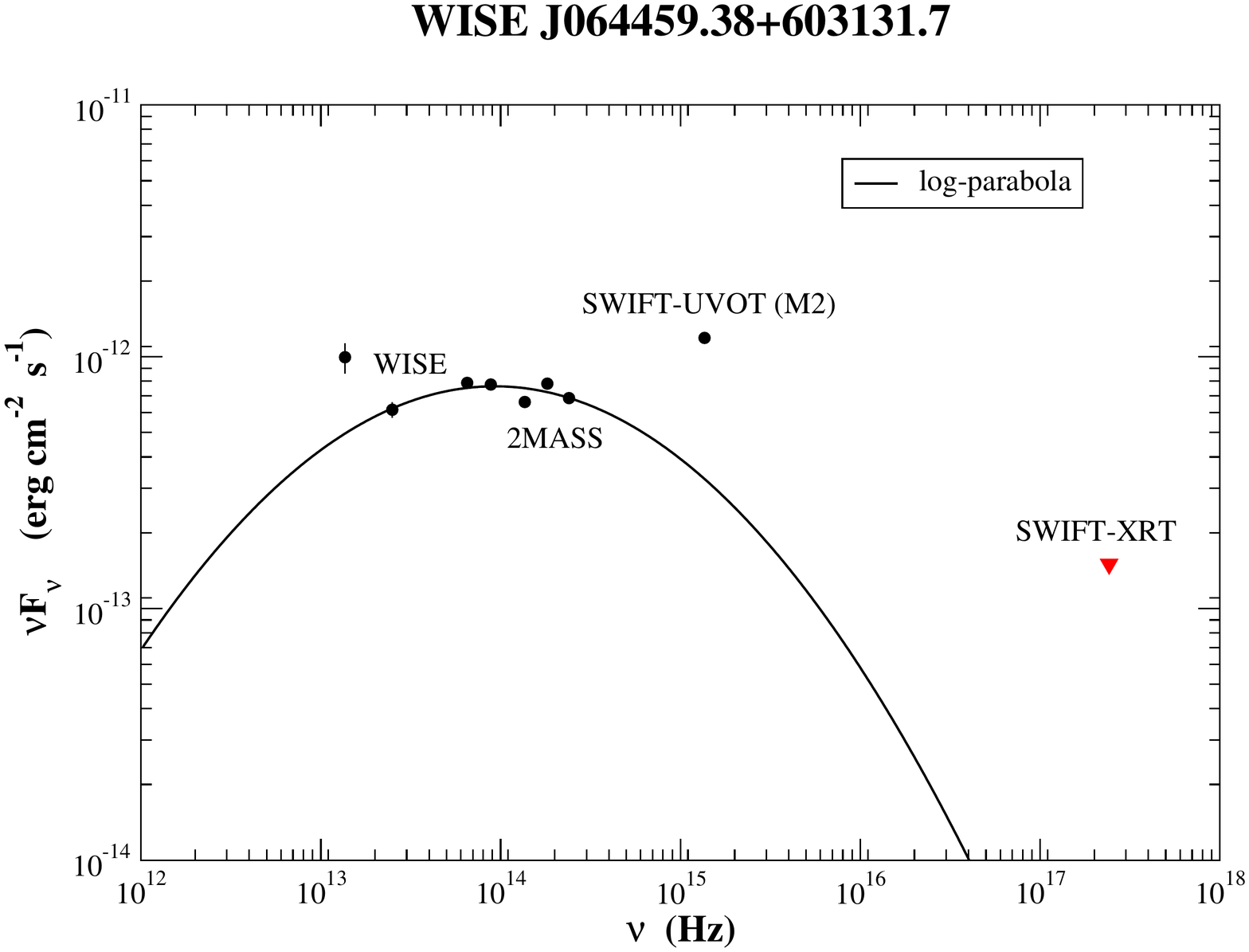}
\includegraphics[height=6.6cm,width=8.8cm,angle=0]{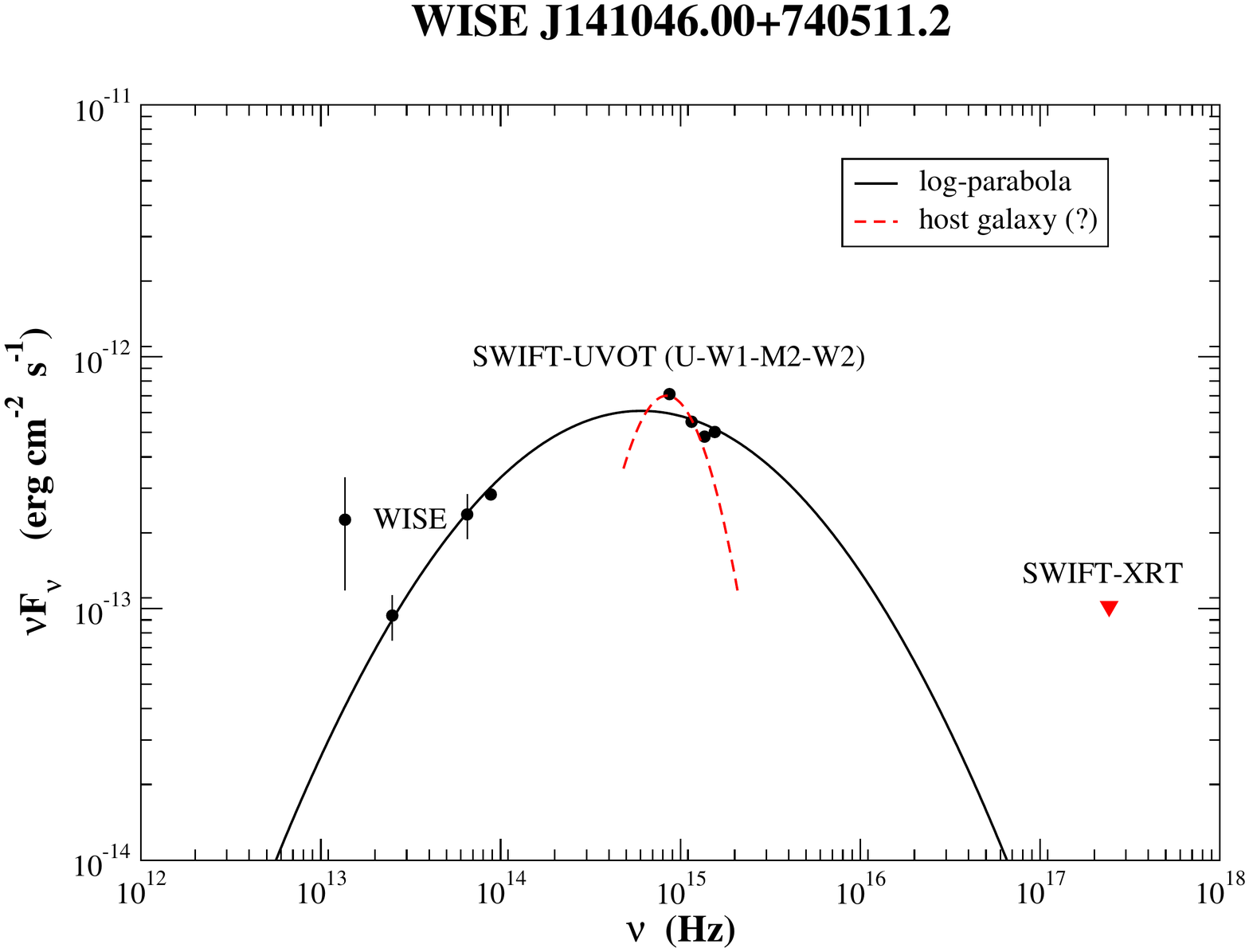}
\caption{The SEDs of both WISE J064459.38+603131 (left panel) and WISE J141046.00+740511.2 (right panel). Data points are shown as black circles and labels indicate the telescope used to obtain them. The black line marks the broad-band log-parabolic function generally adopted to described non-thermal SEDs of BL Lacs. For WISE J141046.00+740511.2 the dashed red line points towards the host galaxy interpretation for the optical-UV data. Unabsorbed X-ray fluxes, in the 0.5-10 keV energy range, estimated from the XRT source counts are also shown as red triangles in both panels.
}
\label{fig:sed}`
\end{figure*}

In Fig.~\ref{fig:sed} we show the SEDs for both WISE J064459.38+603131 and WISE J141046.00+740511.2. For the former SED the infrared emission appear to be compatible with a log-parabolic shape \citep{massaro04,massaro06} with the only exception of the UVOT flux density in the M2 filter. This could be due to the presence of the host galaxy and/or to variability since UV and IR observations were not taken simultaneously and both of them are integrated over different epochs. The latter source: WISE J141046.00+740511.2 has also an SED well described by a log-parabolic shape but in this case the \swf\ optical and UV data seems to mark the presence of an host galaxy.
In both panels of Fig.~\ref{fig:sed} we also report the unabsorbed X-ray flux computed from the source counts of the XRT event file. Since this X-ray flux is an estimate we neglect second order corrections to convert the X-ray flux into a $\nu F_{\nu}$ SED value.

The agreement between the multifrequency data and the broad band log-parabolic SED support the interpretation of WISE J064459.38+603131 and WISE J141046.00+740511.2 as RWBLs.

\section{Summary and Conclusions}
\label{sec:summary} 
\begin{itemize}
\item The mid-IR colors of BL Lacs appear to be clearly different from those of WDs and WELQs. This strongly indicate that the \wse\ color-color diagram built with the mid-IR magnitudes at 3.4$\mu$m, 4.6$\mu$m and 12$\mu$m together with optical spectroscopic observations are a powerful diagnostic tool to distinguish these three source classes. We only found few contaminants within the WD and the WELQ catalogs used in our analysis. However, for all these cases, the combination of mid-IR and optical spectroscopic data allowed us to exclude them as BL Lacs.  
\item Applying this mid-IR color-based analysis we conclude that WISE J064459.38+603131 and WISE J141046.00+740511.2, lying within the positional uncertainty regions of two distinct UGSs, are the first two RWBLs detected so far. Both of them are also associable with 1FHL sources thus strengthening our interpretation, since the high energy sky above 10 GeV is mostly dominated by BZBs. Both sources appear to be variable in the mid-IR as expect for BZBs and they also show a broad-band SEDs, built thanks to the \wse\ and the \swf\ observations, consistent with a log-parabolic shape. 
\item In addition to them 2QZ J215454.3-305654 could also potentially be a RWBL, but lacking of the mid-IR detection at 12$\mu$m it is not possible to confirm its nature using the mid-IR color-color plot. In addition the two contaminants: WISE J001514.87-103043.4 and WISE J150427.69+543902.2, found in the WELQ sample used here, deserve a deeper investigation to clarify their nature.
\end{itemize}

We highlight that the strategy used to search for potential counterparts of UGSs via X-ray follow up observations has been crucial in both cases to discover WISE J064459.38+603131 and WISE J141046.00+740511.2, thus emphasizing that combining X-ray and mid-IR observations could be crucial for future searches of low energy counterparts of gamma-ray sources. However, we also stress that such discovery has been done thanks to optical spectroscopic follow up campaigns that finally revealed the source nature. Since both these two sources are not only associable with 3FGL objects but also with \fer\ sources detected above 10 GeV their existence challenges current gamma-ray association methods mostly based on radio surveys that do not allow to catch these potential counterparts.  The existence of RWBLs will also strong impacts on jet physics since they could allow us to study sources whose emission is dominated by relativistic jets that do not tend to form extended radio structures as their parent population of radio galaxies. 

\begin{figure}[!ht]
\includegraphics[height=8.cm,width=8.8cm,angle=0]{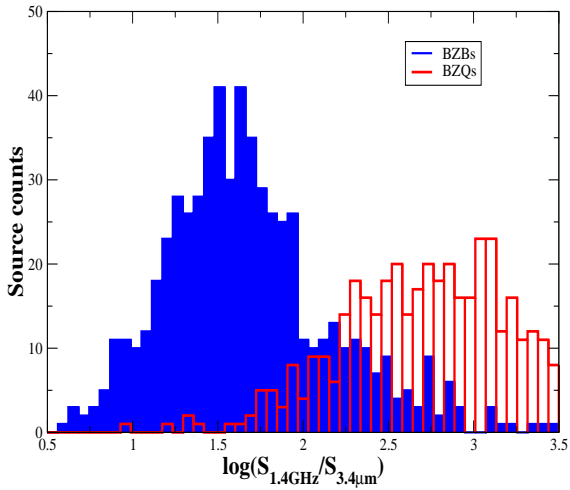}
\caption{The distributions of log of the ratio between the radio flux densities at 1.4 GHz and the mid-IR one at 3.4$\mu$m shown for BZBs (blue) and BZQs (red) in our sample. The two selected RWBLs lie at least on the low tail of the blue distribution drawn for the BZBs thus being extreme sources even if follow up observations will reveal their radio emission.
} 
\label{fig:ratios}
\end{figure}

Finally, we remark that deeper radio follow up observations than those available to date could potentially reveal their low energy emission at GHz frequencies. However the ratio between the infrared and the radio emission of so called RWBLs, being below the flux limit of current radio surveys is unexpectedly unusual. In Fig.~\ref{fig:ratios} we show the distribution of such ratio to highlight how the two selected RWBLs lie in the tail of the BL Lac population being below the 0.5 value.


\acknowledgements
We thank the anonymous referee for useful comments that led to improvements in the paper.
F.M. gratefully acknowledges the financial support of the Programma
Giovani Ricercatori -- Rita Levi Montalcini -- Rientro dei Cervelli (2012) awarded by the Italian Ministry of Education, Universities and Research (MIUR).
H.A.S. acknowledges partial support from NASA Grants NNX15AE56G and NNX14AJ61G.
This research has made use of data obtained from the high-energy Astrophysics Science Archive
Research Center (HEASARC) provided by NASA's Goddard Space Flight Center.
The NASA/IPAC Extragalactic Database
(NED) operated by the Jet Propulsion Laboratory, California
Institute of Technology, under contract with the National Aeronautics and Space Administration.
Part of this work is based on the NVSS (NRAO VLA Sky Survey):
The National Radio Astronomy Observatory is operated by Associated Universities,
Inc., under contract with the National Science Foundation and on the VLA Low-frequency Sky Survey (VLSS).
This publication makes use of data products from the Wide-field Infrared Survey Explorer, 
which is a joint project of the University of California, Los Angeles, and 
the Jet Propulsion Laboratory/California Institute of Technology, 
funded by the National Aeronautics and Space Administration.
TOPCAT\footnote{\underline{http://www.star.bris.ac.uk/$\sim$mbt/topcat/}} 
\citep{taylor05} for the preparation and manipulation of the tabular data and the images.

{}

\end{document}